\documentclass[letter]{aa}
\usepackage{txfonts}
\usepackage{graphicx}

\def\H2{H$_2$}

\begin{document}
\title{First results from the CALYPSO IRAM-PdBI survey\thanks{Based on observations carried out with the IRAM Plateau de Bure interferometer. IRAM is supported by INSU/CNRS (France), MPG (Germany), and IGN (Spain)}}
\subtitle{III. Monopolar jets driven by a proto-binary system in NGC1333-IRAS2A}
\author{C. Codella \inst{1} \and 
A.J. Maury \inst{2,3} \and 
F. Gueth  \inst{4} \and
S. Maret \inst{5} \and 
A. Belloche \inst{6} \and
S. Cabrit \inst{7,5} \and
Ph. Andr\'e \inst{8}}

\institute{
INAF, Osservatorio Astrofisico di Arcetri, Largo E. Fermi 5,
50125 Firenze, Italy
\and
Harvard-Smithsonian Center for Astrophysics, 60 Garden street, Cambridge, MA 02138, USA
\and
ESO, Karl Schwarzchild Str. 2, 85748 Garching bei M\"unchen, Germany
\and
IRAM, 300 rue de la Piscine, 38406 Saint Martin d'H\`eres, France
\and
UJF-Grenoble1/CNRS-INSU, Institut de
Plan\'etologie et d'Astrophysique de Grenoble (IPAG) UMR 5274,
Grenoble, 38041, France
\and
Max-Planck-Institut f\"ur Radioastronomie, Auf dem H\"ugel 69, 53121 Bonn, Germany
\and
LERMA, Observatoire de Paris, CNRS, ENS, UPMC, UCP, 61 Av. de l'Observatoire, 75014 Paris, France
\and
Laboratoire AIM-Paris-Saclay,  
CEA/DSM/Irfu - CNRS -Universit\'e 
Paris Diderot, CE Saclay, 91191 Gif-sur-Yvette Cedex, France
}

\offprints{C. Codella, \email{codella@arcetri.astro.it}}
\date{Received date; accepted date}

\authorrunning{Codella et al.}
\titlerunning{Monopolar jets driven by a proto-binary system in NGC1333-IRAS2A}

\abstract
{The earliest evolutionary stages of low-mass protostars are characterised by hot and fast jets which remove 
angular momentum from the circumstellar disk, thus allowing mass accretion onto the central object. 
However, the launch mechanism is still being debated.}
{We would like to exploit high-angular ($\sim$ 0$\farcs$8) resolution and high-sensitivity images to investigate  
the origin of protostellar jets using typical molecular tracers of shocked regions, such as SiO and SO.}
{We mapped the inner 22$\arcsec$ of the NGC1333--IRAS2A protostar in SiO(5--4), SO(6$_5$--5$_4$), and the continuum
emission at 1.4 mm using the IRAM Plateau de Bure interferometer in the framework of the CALYPSO IRAM large program.}
{For the first time, we disentangle the NGC1333--IRAS2A Class 0 object into 
a proto-binary system revealing two protostars (MM1, MM2) separated by $\sim$ 560 AU, each of them driving their own jet, 
while past work considered a single protostar with a quadrupolar outflow.
We reveal (i) a clumpy, fast (up to $\vert$$V$--$V_{\rm LSR}$$\vert$ $\geq$ 50 km s$^{-1}$), and 
blueshifted jet emerging from the brightest MM1 source, and (ii) 
a slower redshifted jet, driven by MM2.  
Silicon monoxide emission is a powerful tracer of high-excitation ($T_{\rm kin}$ $\geq$ 100 K; $n_{\rm H_2}$ $\geq$ 10$^5$ cm$^{-3}$) 
jets close to the launching region.
At the highest velocities, SO appears to mimic SiO tracing the jets,
whereas at velocities close to the systemic one, SO is dominated by extended emission,   
tracing the cavity opened by the jet.} 
{Both jets are intrinsically monopolar, and intermittent in time. 
The dynamical time of the SiO clumps is $\leq$ 30--90 yr, 
indicating that one-sided ejections from protostars can take place on these timescales.}
 

\maketitle

\section{Introduction}

The so-called Class 0 objects represent the earliest low-mass protostellar
stage having (i) most of their
mass still in the form of dense envelopes,
and (ii) a lifetime $\leq$ a few 10$^5$ yr (e.g. Andr\'e et al. 2000; Evans et al. 2009; Maury et al. 2011).
Class 0 protostars then represent an ideal laboratory for tracing the 
pristine conditions of low-mass star formation. Because of the paucity of the sub-arcsec
(sub)mm observations required to probe the innermost ($\le$ 100 AU) regions, several basic questions
remain open, such as the existence of multiple systems, or the launching
mechanism of protostellar jets. 
Protostars drive fast jets surrounded by wide-angle winds that impact 
the high-density parent cloud generating shock fronts, which trigger endothermic  
chemical reactions and ice grain mantle sublimation or sputtering. 
As a consequence, several molecules (such as H$_2$O, CH$_3$OH, and S-bearing species) undergo
significant enhancements in their abundances (e.g. van Dishoeck
\& Blake 1998). A typical example is represented by SiO, whose formation
is mainly ($\ge$ 90\%) attributed to the sputtering of Si atoms from
refractory core grains in high-velocity ($\ge$ 20 km s$^{-1}$) shocks 
(e.g. Gusdorf et al. 2008ab), or grain shattering in grain-grain collisions inside J-shocks (Guillet et al. 2010).
Silicon monoxide traces shocks inside jets well, suffering minimal
contamination from low-velocity swept-up material
(usually traced by low-J CO emission), and is able
to unambiguously probe the mass loss process.  

So far, a quite limited number of Class 0 jets has been observed 
at sub-arcsecond angular resolution 
(needed to disentangle the jet and the outflow cavities):
HH211 (Lee et al. 2007, 2009, 2010), HH212 (Codella et al. 2007,
Lee et al. 2008), 
IRAS04166+2706 (Tafalla et al. 2010),
and L1448-C (Maury et al. 2010, Hirano et al. 2010).
The IRAM Plateau de Bure interferometer (PdBI) large program CALYPSO\footnote{http://irfu.cea.fr/Projects/Calypso} 
(Continuum and Lines from Young ProtoStellar Objects) 
is correcting this situation by providing the 
first sub-arcsecond statistical study of inner jet properties in 
nearby low-luminosity Class 0 sources in combination with studies of the 
envelopes, disks, and multiplicity structure. 
One of the best documented CALYPSO targets is NGC1333-IRAS2A (hereafter IRAS2A), located
at 235 pc\footnote{Recent estimates of the distance to Perseus range from 220 to
350 pc. Here we adopt 235 pc following Hirota et al. (2008).} in the Perseus NGC1333 cluster.
The source IRAS2A is part of a wider system containing IRAS2B (not investigated here), located 
at $\sim$ 31$\arcsec$. The IRAS2A luminosity is $\sim$ 10 $L_{\rm \sun}$, 
and it was observed 
in continuum at cm (e.g. Reipurth et al. 2002),
mm (Looney et al. 2000; J\o{}rgensen et al. 2004a, 2007, 2009; Maury et al. 2010), 
and sub-mm wavelengths (e.g. Sandell \& Knee 2001). 
The outflow activity was traced using single-dish telescopes and interferometers and several
tracers of swept-up material (e.g. CO) and shocks
(e.g. SiO, CH$_3$OH), revealing two perpendicular outflows, directed NE-SW
(PA $\simeq$ 25$\degr$; hereafter called N-S for sake of clarity) and SE-NW (PA $\simeq$ 105$\degr$;
hereafter E-W), both originating to within a few arcseconds from IRAS2A  
(e.g. Bachiller et al. 1998; Knee \& Sandell 2000; 
J\o{}rgensen et al. 2004ab, 2009; Wakelam et al. 2005; Persson et al. 2012; Plunkett et al. 2013).
These outflows seem intrinsically different, the E-W outflow being more 
collimated and chemically richer than the N-S one, 
supporting the possibility that IRAS2A is an  
unresolved proto-binary. 

\section{Observations} 

The source IRAS2A was observed with the IRAM PdB six-element array in December 2010 and January-February 2011 
using both the A and C configurations.  
The shortest and longest baselines are 19 m and 762 m, respectively, allowing us to recover  
emission at scales from $\sim$ 8$\arcsec$ down to 0$\farcs$4 at 1.4 mm.
The SiO(5--4) and SO(6$_5$--5$_4$) lines\footnote{Spectroscopic parameters have been extracted from the Jet
Propulsion Laboratory molecular database (Pickett et al. 1998).}  at 217104.98 
and 219949.44 MHz, 
respectively, were observed  
using the WideX backend to cover a 4 GHz spectral window 
and to probe continuum emission  
at a 2 MHz ($\sim$ 2.6 km s$^{-1}$ at 1.4 mm) spectral resolution.
Calibration was carried out following standard procedures, 
using GILDAS-CLIC\footnote{http://www.iram.fr/IRAMFR/GILDAS}.
Phase (rms) was $\le$ 50$\degr$ and 80$\degr$ for the A and C tracks, respectively, 
pwv was 0.5–-1 mm (A) and $\sim$ 1--2 mm (C), and system temperatures were 
$\sim$ 100--160 K (A) and 150--250 K (C).
The final uncertainty on the absolute flux scale is $\leq$ 15\%. 
The typical rms noise in the 2 MHz channels was 3--9 mJy beam$^{-1}$. 
Images were produced using robust weighting, and restored with a clean beam of
$0\farcs81\times0\farcs69$ (PA = 33$\degr$). 

\begin{figure*}
\centerline{\includegraphics[angle=0,width=15.5cm]{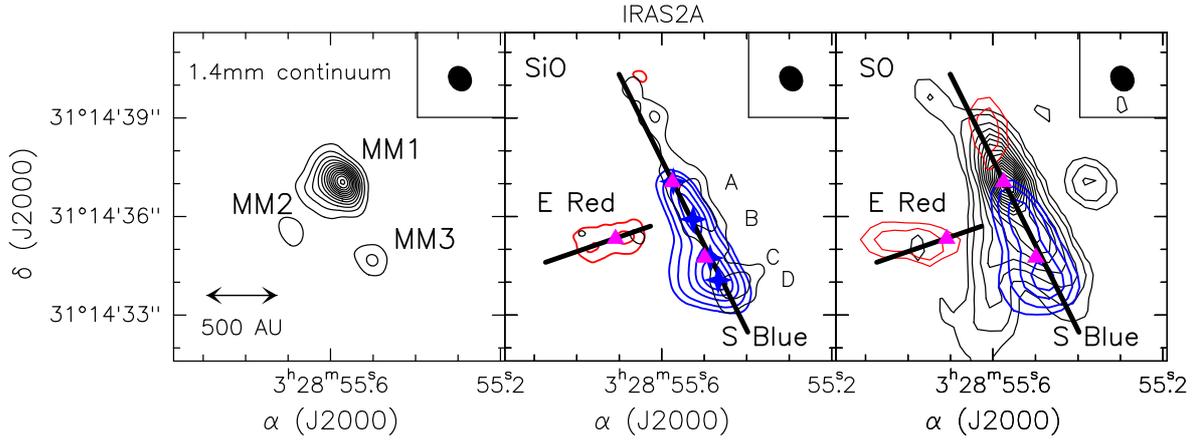}}
\caption{{\it Left panel:} Contour plots of the IRAS2A continuum emission at 
1.4 mm. The ellipse shows
the PdBI synthesised beam (HPBW): $0\farcs81\times0\farcs69$ (PA = 33$\degr$). 
First contours and steps correspond to 5$\sigma$ (7.5 mJy beam $^{-1}$).
Labels indicate the main source (MM1) and two weaker sources (MM2 and MM3). 
{\it Middle panel:}
Contour map of blue- (--39,+3 km s$^{-1}$) and redshifted (+11,+21 km s$^{-1}$) SiO(5--4) emissions
superimposed on the SiO at low-velocity (+3,+11 km s$^{-1}$; black contours). 
First contours correspond 
to 5$\sigma$ and 10$\sigma$, followed by steps of 10$\sigma$.
One $\sigma$ is 84 (blue) , 15 (red), and 12 (black) mJy beam $^{-1}$ km s$^{-1}$.
Crosses are for the position of the four SiO clumps 
(A, B, C, and D; see Figs. 2 and B.3).
Magenta triangles stand for the positions of MM1, MM2, and MM3.
Grey lines are the directions of the PV diagrams shown in Figs. 2 and B.3.
{\it Right panel:} Same as Middle panel for the SO(6$_5$-5$_4$), 
averaged over (--29,+3), 
(+11,+19), and (+3,+11) km s$^{-1}$ for the blue-, red-, and black-velocity, respectively.
One $\sigma$ is 86 (blue) , 18 (red), and 16 (low-velocity) mJy beam $^{-1}$ km s$^{-1}$.}
\label{maps}
\end{figure*}

\section{Results and discussion}

\subsection{Continuum emission}

\begin{table}
\center
\caption[] {Position and intensity of the continuum peaks.}
\label{table_cont}
\begin{tabular}{lccc}
\hline
\multicolumn{1}{c}{Source} &
\multicolumn{1}{c}{$\alpha({\rm J2000})$$^a$} &
\multicolumn{1}{c}{$\delta({\rm J2000})$$^a$} &
\multicolumn{1}{c}{$I^{\rm peak}_{\rm 1.4mm}$} \\ 
\multicolumn{1}{c}{ } &
\multicolumn{1}{c}{(03$^h$ 28$^m$ $^s$)}&
\multicolumn{1}{c}{(+31$\degr$ 14$\arcmin$ $\arcsec$)} &
\multicolumn{1}{c}{(mJy beam$^{-1}$)} \\ 
\hline
MM1 & 55.58 & 37.06 & 94(2)  \\
MM2 & 55.71 & 35.33 & 15(1)  \\
MM3 & 55.50 & 34.75 & 21(1)  \\
\hline
\end{tabular}
\begin{center}
$^a$ The fit uncertainties are
4, 14, and 17 mas for MM1, MM2, and MM3, respectively.
\end{center}
\end{table}

Emission map of the 1.4 mm continuum
is shown in Fig.~1.
The source IRAS2A is found to be associated with three continuum sources
(here labelled MM1, MM2, and MM3).
A detailed analysis of the continuum emission
is beyond the scope of the present paper: it will be used to
support the interpretation of the SiO and SO images.
Table 1 summarises positions and 1.4 mm peak fluxes 
of the three continuum sources. The coordinates of 
the brightest one (MM1) are 
consistent with the position of IRAS2A previously
measured using the VLA (3.6 cm), SMA (0.8 and 1.3 mm), and BIMA 
(2.7 mm) telescopes (Rodr\'{\i}guez et al. 1999; 
J\o{}rgensen et al. 2007; Looney et al. 2007). 
In addition, a fainter and spatially unresolved source (MM2)
is found $\sim$ 2$\farcs$4 (560 AU) from MM1 in the SE direction. 
Both MM1 and MM2 have also been detected at 94 GHz in the
framework of CALYPSO (see Appendix A): the spectral 
index $\alpha$ (where flux density $S_{\rm \nu}$ $\propto$ $\nu^\alpha$) is $\sim$ 2-2.5, consistent with that of a protostar.

A third source (MM3) is detected $\sim$ 2$\farcs$5 south of MM1.
Its FWHP size is 307 mas and its non-detection at 94 GHz 
(with a peak flux $\leq$ 0.1 mJy beam$^{-1}$ implying $\alpha$ $\geq$ 4) 
challenges a protostellar nature. 
Alternatively, MM3 might be an outflow feature due to 
dust heated by shocks travelling along the SiO jet (see Sect. 3.2). 

\subsection{Different jets from a proto-binary system}

Figure 1 shows that SiO(5--4) emission  
is mainly confined to a collimated blueshifted southern SiO jet with a PA of 25$\degr$,   
emerging from MM1, and extending out to $\sim$ 4$\arcsec$ (1000 AU). 
The SiO jet is narrow: after correction for the PdBI HPBW, the transverse 
FWHM is $\simeq$ 0$\farcs$7$\pm$0$\farcs$1 (165 AU)  at $\sim$ 700 AU from MM1, while it appears even narrower  
(being spatially unresolved) close to the driving source. 
Position-velocity (PV) diagrams along the N-S jet axis (Fig. 2) 
show that SiO emission extends to very high blueshifted velocities,
$\sim$ --50 km s$^{-1}$ with respect to 
$V_{\rm LSR}$\footnote{The $V_{\rm LSR}$ of IRAS2A as given in the literature  
lies between +7.0 km s$^{-1}$ and +7.7 km s$^{-1}$ (e.g. Persson et al. 2012, and references therein);
we adopt +6.5 km s$^{-1}$, according
to CALYPSO measurements of high-excitation ($\sim$ 200 K) hot-core tracers, 
Maret et al. (A\&A, in press).}
= +6.5 km s$^{-1}$.

The MM1 SiO jet is surprisingly asymmetric with a bright (up to 90 K in $T_{\rm MB}$ scale, see e.g. Fig. 3) 
blueshifted emission and no clear
red counterpart (down to 1 K), suggesting a monopolar nature. 
The presence of monopolar outflows has recently been observed by Fern\'andez-L\'opez et al. (2013)
towards the complex high-mass star forming region IRAS18162-2048. In that case, the authors propose
precession and deflection due to high-density clumps to explain the asymmetric appearance.
In principle, asymmetries in ambient gas could affect emission at low velocities (such as
swept-up gas, see e.g. Pety et al. 2006), but not the jet emission.
As far as we know, this is the first time a SiO monopolar high-velocity jet                 
ejected from a low-mass protostar has been observed. 
The lack of SiO redshifted emission could be due to the lack of dust if the 
northern cavity has been completely evacuated by previous ejections. 
However, the lack the high-velocity redshifted emission in SO (see Sect. 3.3),
whose abundance increases due to pure gas phase neutral-neutral reactions, seems
to rule out this hypothesis.
As a consequence, the bright blueshifted jet from MM1 argues that,  
intrinsically, one-sided ejections from low-mass protostars can occur,
i.e. that one side of the accreting disk is ejecting more material than the other.
A N-S outflow on a large scale ($\sim$ 2$\arcmin$) was previously detected
with both single-dish antennas and interferometers using CO(1--0) and (2--1) 
(e.g. Engargiola \& Plambeck 1999), showing extended lobes at
relatively low velocity ($\vert$$V$--$V_{\rm LSR}$$\vert$ $\leq$ 10 km s$^{-1}$).
Bipolar non-collimated N-S emission has been also traced 
on  10$\arcsec$--20$\arcsec$ angular scales 
using CS, HCO$^+$, and HCN emission at even lower velocities 
($\vert$$V$--$V_{\rm LSR}$$\vert$ $\leq$ 5 km s$^{-1}$; J\o{}rgensen et al. 2007, 2009).
Maret et al. (2009) observed bipolar H$_2$ emission using the Spitzer telescope. 
Therefore, the present SiO image reveals
for the first time the fast jet sweeping up the slower outflow observed on larger scale.
The jet kinematical age, derived from the farthest SiO emission, is 88 years. 
Given that the jet maps suggest an inclination $\theta$ with respect to the
plane of the sky $\leq$ 45$\degr$, this estimate has to be
considered an upper limit\footnote{The age should be corrected by a factor of ctg($\theta$).}.
In conclusion, given the bipolarity of CO on large scales, the N-S ejection was symmetric in the past, whereas
the present SiO image suggests that in the last $\sim$ 90 years only the southern side has been active.

Four distinct clumps (labelled A, B, C, and D), to first-order tracing a sequence of
shocks along the jet, are  
clearly visible at different velocities and different positions along the 
bright SW blue lobe; their offsets with respect to MM1 are:
(--0$\farcs$01,+0$\farcs$02), (--0$\farcs$59,--1$\farcs$13), (--0$\farcs$86,--2$\farcs$31), and
(--1$\farcs$35,--2$\farcs$90), respectively. 
Clump A, emitting at the highest velocities, is closely associated with MM1, confirming that SiO  
is a powerful tracer of the jet at the base in Class 0 sources (e.g.
Codella et al. 2007). Clump C (peaking at $\sim$ --25 km s$^{-1}$)  
is instead associated with the MM3 continuum source; 
MM3 could be a young stellar object driving the 
blueshifted SiO emission (between --10 and 0 km s$^{-1}$,
as shown in Fig. 2; see also the channel maps reported in Fig. B.1)
which deviates from the N-S
main axis, bending towards the east. 
Alternatively, the continuum source MM3 could trace dust emission from
a pre-existing clumpy denser region which, as a side effect, bends part of the blue flow. 
Finally, the PV diagram of Fig. 2 suggests a jet deceleration. 
The dynamical time of the SiO clumps is $\leq$ 27--88 yr, and is consistent with that 
derived for the HH212 SiO jet (25 yr; Cabrit et al. 2007).

In addition to the N-S jet, the SiO map reveals a redshifted jet 
($V$--$V_{\rm LSR}$ up to $\sim$ +12 km s$^{-1}$; see Fig. B.1)  
with a width similar to the N-S jet (165 AU) and spatially associated with the MM2 continuum source,
confirming SiO as a probe of the jet launching region. 
The jet is monopolar in this case as well 
and seems to decelerate (see the channel maps and the PV diagrams in Figs. B.1 and B.3).
The elongation of the jet is
consistent with the PA ($\sim$ 105$\degr$) of the E-W outflow, which consists of
two highly collimated lobes observed quite 
far (60$\arcsec$--80$\arcsec$) from IRAS2A, 
using typical tracers  
(such as SiO, SO, SO$_2$, and CH$_3$OH) of shock chemistry
(e.g. Bachiller et al. 1998; Wakelam et al. 2005).  
So far, the driving sources of the two perpendicular E-W and N-S outflows  
have not been revealed. The present SiO (and continuum) images allow us to 
resolve for the first time the origin of the IRAS2A quadrupolar outflow, 
unveiling a Class 0 proto-binary system (MM1 and MM2) driving two different jets.

\subsection{The role of SO emission: jets and cavities}

\begin{figure}
\centerline{\includegraphics[angle=0,width=6.5cm]{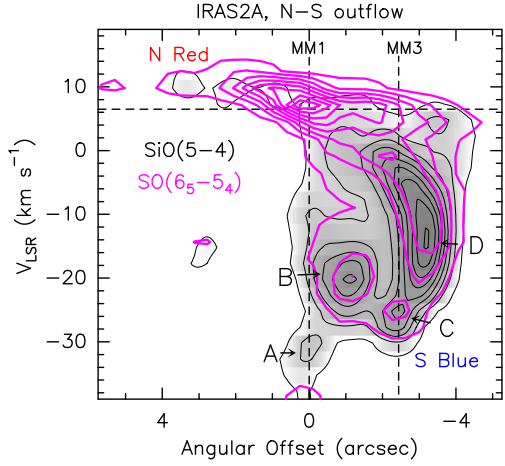}}
\caption{Position-velocity cut of SiO(5--4) (grey scale and black contours)
and SO(6$_5$--5$_4$) (magenta contours) along the
N-S jet (PA = 25$\degr$, see grey line in Fig. 1).
First contours and steps correspond to 5$\sigma$ (3.0 K for SiO and 4.5 K for SO)
and 10$\sigma$, respectively. Dashed lines mark the positions of MM1, MM3, and the 
ambient $V_{\rm LSR}$ (+6.5 km s$^{-1}$). Labels A, B, C, and D are for
the four clumps along the SiO blue jet. No SiO or SO emission
is detected outside the given velocity range.}
\label{pvsioso}
\end{figure}

At the highest velocities, the SO distribution, as traced by its (6$_5$--5$_4$) line (Fig. 1), resembles
the SiO(5--4) one, showing a bright S jet driven by MM1, and supporting the association with the SiO jet itself.
Figure 3 plots as an example the SiO and SO spectra observed towards clump C, 
confirming that they are very similar at the highest velocities. 
These findings (i) are in agreement with the detection of SO at 
extremely high velocities ($\vert$$V$--$V_{\rm LSR}$$\vert$ $\geq$ 50 km s$^{-1}$) using the
IRAM 30 m antenna towards the L1448 and IRAS04166+2706 outflows (Tafalla et al. 2010),
and (ii) confirm what was found by Lee et al. (2010) for HH211, i.e.  
that SO can be used as molecular jet tracer in addition to the  
well-known H$_2$, CO, and SiO
(and H$_2$O masers), bringing a new constraint on jet chemical models. 
Indeed, magnetohydrodynamic (MHD) models show that  
the SO abundance, quickly formed by the reaction of S with OH, can reach the observed abundance
of 2 10$^{-7}$ in jets (Tafalla et al. 2010) through ambipolar diffusion heating
in C-shocks (Pineau Des For\^ets et al. 1993) or magneto-centrifugal disk winds (Panoglou et al. 2012).

Close to the systemic velocity, 
the SiO intensity fades whereas SO increases. 
This is particularly clear when we compare the 
profiles observed towards clump C (Fig. 3) and the spatial
distributions in Fig. 1:
low-velocity ($\vert$$V$--$V_{\rm LSR}$$\vert$ $\le$ 4 km s$^{-1}$) SO bright emission traces  
extended emission in both the northern and southern lobes  
(Fig. 2; see also the channel maps of Fig. B.2).  
In particular, Fig. 1 suggests the association of low-velocity SO with a cavity with MM1 at the vertex. 
Emission of SO redshifted by $\sim$ 5 km s$^{-1}$ is also detected towards north
in addition to the SiO MM2 jet, but the
morphology suggests that this emission is still associated with a cavity rather than the jet.
In addition, an SO eastern clump redshifted by 2--3 km s$^{-1}$ appears along the direction of the E-W jet, 
and is plausibly related to swept-up material.
The low-velocity SiO emission
is elongated, but it is definitely weaker and offset to the NW
with respect to the blue jet axis, and supports its association with the SO cavity.
The weakness of SiO in the cavity should reflect its low formation rate in
low-velocity shocks (e.g. Gusdorf et al. 2008ab).
Interestingly, the H$_2^{18}$O emission imaged at PdBI by Persson et al. (2012) and distributed along the
direction of the blueshifted outflow, is emitting in the +1,+9 km s$^{-1}$ range, suggesting
that H$_2^{18}$O also traces the outflow cavities.   
In summary, the low-velocity emission traces a cavity opened by the fast jet, as predicted by MHD disk wind 
models (Cabrit et al. 1999).

\subsection{High brightness temperatures and excitation conditions}

The SiO(5--4) profiles reveal extremely high
brightness temperatures $T_{\rm MB}$ of up to 90 K. 
These values are compared with the result of
the RADEX\footnote{http://home.strw.leidenuniv.nl/$\sim$moldata/radex.html}
non-LTE code (van der Tak et al. 2007) with the
rate coefficients for collisions with H$_2$
(Dayou \& Balan\c{c}a 2006) using a plane
parallel geometry, and assuming a FWHM linewidth of 20 km s$^{-1}$.
One line is obviously not enough for a proper analysis; nevertheless,
if we assume $T_{\rm kin}$ $\leq$ 500 K, the high $T_{\rm MB}$ values constrain the total SiO column densities
$N_{\rm SiO}$ $\geq$ 10$^{15}$ cm$^{-2}$. 
Interestingly, the highest $T_{\rm MB}$ suggests
high excitation conditions with $T_{\rm kin}$ $\geq$ 100 K and
$n_{\rm H_2}$ $\geq$ 10$^5$ cm$^{-3}$, in agreement with the estimates found
for SiO clumps associated with other protostellar outflows (e.g.
Hirano et al. 2006; Nisini et al. 2007; Cabrit et al. 2007),  confirming the association
of SiO with shocked material.
If we model the $T_{\rm MB}$ $\sim$ 30 K of the high-velocity SO(6$_5$--5$_4$) emission
observed towards clump C using RADEX coupled with the collision rates provided by Green (1994),
we find $N_{\rm SO}$ $\sim$ 10$^{16}$--10$^{17}$ cm$^{-2}$ 
and $n_{\rm H_2}$ $\geq$ 10$^5$ cm$^{-3}$, supporting, as for SiO, shocked (compressed) gas.

\begin{figure}
\centerline{\includegraphics[angle=0,width=7cm]{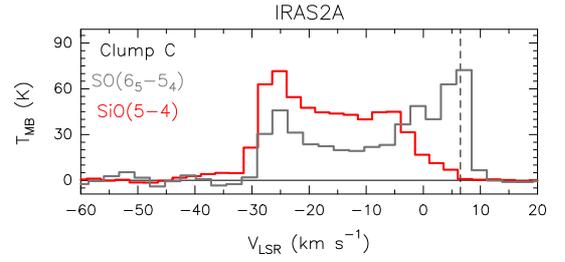}}
\caption{Comparison between the SiO(5--4) and SO(6$_5$--5$_4$)
lines as observed towards clump C (in main-beam temperature, $T_{\rm MB}$, scale).}
\label{spectra}
\end{figure}

\section{Conclusions}

The present continuum, SiO, and SO data allow us to disentangle the origin
of the IRAS2A quadrupolar outflow   
into a proto-binary system powering two different jets.
We revealed a clumpy S 
jet emerging from the brightest MM1 continuum source, plus  
a redshifted E jet associated with the weaker MM2 source.
The jet gas has high-excitation conditions ($\geq$ 100 K; $\geq$ 10$^5$ cm$^{-3}$).
The fast, young ($\leq$ 90 yr) S jet opened a molecular cavity,  
efficiently traced by SO at velocity close to systemic ($\vert$$V$--$V_{\rm LSR}$$\vert$ $\leq$ 4 km s$^{-1}$). 
The IRAS2A jets are intrinsically monopolar on scales $<$ 1000 AU indicating
that one-side ejections from protostars are possible during short periods ($\leq$ 90 yr).

\begin{acknowledgements}
We are very grateful to all the IRAM staff, whose dedication allowed us to carry out the CALYPSO project.
The research leading to these results has received funding from the European Community’s Seventh 
Framework Programme (FP7/2007-2013/) under grant agreements No 229517 (ESO COFUND) and No 291294 (ORISTARS),
and from the French Agence Nationale de la Recherche (ANR), under reference ANR-12-JS05-0005.
\end{acknowledgements}

\vspace{0.5cm}
\noindent
{\bf References} \\

\noindent
Andr\'e Ph., Ward-Thompson D., \& Barsony M. 2000, Protostars and Planets IV, 59 \\
\noindent
Bachiller R., Codella C., Colomer F., Liechti S., \& Walmsley C.M. 1998, A\&A 335, 266  \\
\noindent
Cabrit S., Ferreira J., \& Raga A.C. 2007, A\&A 343, L61  \\
\noindent
Cabrit S., Codella C., Gueth F., et al. 2007, A\&A 468, L29  \\
\noindent
Codella C., Cabrit S., Gueth F., et al. 2007, A\&A 462, L53  \\
\noindent
Dayou F., \& Balan\c{c}a C. 2006, A\&A 459, 297 \\
\noindent
Engargiola G., \& Plambeck R.L. 1999, The Physiscs and Chemistry of the Interstellar Medium, 
Verlag, 291 \\
\noindent
Evans N.J., Dunham M.M., J\o{}rgensen J.K., et al. 2009, ApJS 181, 321 \\
\noindent
Fern\'andez-L\'opez M., Girart J.M., Curiel S., et al. 2013, ApJ 778, 72 \\
\noindent
Green S. 1994, ApJ 434, 188 \\
\noindent
Guillet V., Jones A.P., \& Pineau Des For\^ets G.
2010, A\&A 497, 145  \\
\noindent
Gusdorf A., Cabrit S., Flower D.R., \& Pineau Des For\^ets G.
2008a, A\&A 482, 809 \\
\noindent
Gusdorf A., Pineau Des For\^ets G., Cabrit S., \& Flower D.R.
2008b, A\&A 490, 695 \\
\noindent
Hirano N., Liu S.-Y., Ho P.P.T., et al. 2006, ApJ 636, L141  \\
\noindent
Hirano N., Ho P.P.T., Liu S.-Y., et al. 2010, ApJ 717, 58 \\
\noindent
Hirota T., Bushimata T., Choi Y.K., et al. 2008, PASJ 60, 37 \\
\noindent
J\o{}rgensen J.K., Hogerheijde M.R., van Dishoeck E.F., Blake G.A., \& Sch\"{o}ier F.L.
2004a, A\&A 413, 993 \\
\noindent
J\o{}rgensen J.K., Hogerheijde M.R., \& Blake G.A. 2004b, A\&A 415, 1021 \\
\noindent
J\o{}rgensen J.K., Bourke T.L., Myers P.C., et al. 2007, ApJ 659, 479 \\
\noindent
J\o{}rgensen J.K., van Dishoeck E.F., Visser R., et al. 2009, A\&A 507, 861  \\
\noindent
Knee L.B.G., \& Sandell G. 2000, A\&A 361, 671 \\ 
\noindent
Lee C.-F., Ho P.T.P., Hirano N., et al. 2007, ApJ 659, L499 \\
\noindent
Lee C.-F., Ho P.T.P., Bourke T.L., et al. 2008, ApJ 685, L1026 \\
\noindent
Lee C.-F., Hirano N., Palau A., et al. 2009, ApJ 699, L1584 \\
\noindent
Lee C.-F., Hasegawa T.I., Hirano N., et al. 2010, ApJ 713, L731 \\
\noindent
Looney L.W., Mundy L.G., \& Welch W.J 2000, ApJ 529, L477 \\
\noindent
Maret S., Bergin E.A., Neufeld D.A., et al. 2009, A\&A 698, 1244 \\ 
\noindent
Maury A.J., Andr\'e Ph., Hennebelle, P., et al. 2010 A\&A 512, 40 \\
\noindent
Maury A.J., Andr\'e Ph., Men'shchikov A., K\"onyves V., \& Bontemps S. 2011, A\&A 535, 77  \\
\noindent
Nisini B., Codella C., Giannini T., et al. 2007 A\&A 462, 163 \\
\noindent
Panoglou D., Cabrit S., Pineau Des For\^ets G., et al. 2012, A\&A 538, A2  \\
\noindent
Persson M.V, J\o{}rgensen J.K., \& van Dishoeck E.F. 2012, A\&A 541, A39 \\
\noindent
Pety J., Gueth F., Guilloteau S., \& Dutrey A. 1996, A\&A 458, 841 \\
\noindent
Pickett H.M., Poynter R.L., Cohen E.A.,  et al. 1998,
J. Quant. Spectrosc. \& Rad. Transfer 60, 883 \\
\noindent
Pineau des For$\hat {\rm e}$ts G., Roueff E., Shilke P., \& Flower D.R. 1993, MNRAS 262, 915 \\
\noindent
Plunkett A.L., Arce H.G., Corder S.A., et al. 2013, ApJ 774, 22 \\
\noindent
Reipurth B., Rodr\'{\i}guez L., Anglada G., \& Bally J. 2002, AJ 124, 1045 \\
\noindent
Sandell G., \& Knee L.B.G. 2001, ApJ 546, L49 \\
\noindent
Tafalla M., Santiago-Garc\'{\i}a J., Tafalla M., Johnstone D., \& Bachiller R. 2009, A\&A 465, 169  \\
\noindent
Tafalla M., Santiago-Garc\'{\i}a J., Hacar A., \& Bachiller R. 2010, A\&A 522, 91 \\
\noindent
van der Tak F.F.S., Black J.H., Sch{\o}ier F.L., et al. 2007, A\&A 468, 627 \\
\noindent
van Dishoeck E.F., \& Blake G.A., 1998, ARA\&A 36, 317 \\
\noindent
Wakelam, V., Ceccarelli C., Castets A., et al. 2005, A\&A 437, 149 \\

\clearpage 

\appendix

\section{The 3.2 mm continuum emission}

Figure A.1 shows the emission map of the 3.2 mm continuum dust emission,
which was produced as the 1.4 mm map using robust weighting, and restored with a clean beam of
$1\farcs42\times1\farcs00$ (PA = 38$\degr$).
The 3.2 mm emission allows us to detect the MM1 
($\alpha({\rm J2000})$: 03$^h$ 28$^m$ 55$\fs$56, $\delta({\rm J2000})$: +31$\degr$ 14$\arcmin$ 36$\farcs$93) and MM2 
($\alpha({\rm J2000})$: 03$^h$ 28$^m$ 55$\fs$69, $\delta({\rm J2000})$: +31$\degr$ 14$\arcmin$ 35$\farcs$63) sources, consistent with what was found in 
the 1.4 mm image (see Table 1 and Fig. 1).
The peak fluxes are 17 mJy beam$^{-1}$ and 2 mJy beam$^{-1}$ for MM1 and MM2, respectively.
On the other hand, MM3 (revealed at 1.4 mm) is not detected at a 3$\sigma$ sensitivity level of 0.75 mJy beam$^{-1}$.

\begin{figure}
\centerline{\includegraphics[angle=0,width=7.5cm]{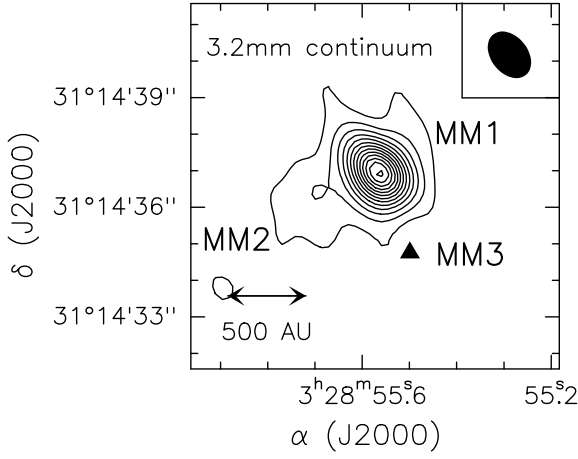}}
\caption{Contour plots of the IRAS2A continuum emission at
3.2 mm. The ellipse shows
the PdBI synthesised beam (HPBW): $1\farcs42\times1\farcs00$ (PA = 38$\degr$).
First contours and steps correspond to 5$\sigma$ (1.3 mJy beam $^{-1}$).
Labels indicate the main source MM1 and the weaker source MM2.
The black triangle stands for the position of MM3, revealed at 1.4 mm and not detected at 3.2 mm.}
\label{3mmmap}
\end{figure}

\section{SiO and SO channel maps}

We show in Figs. A.1 and A.2 the channel maps of the SiO(5--4) 
and SO(6$_5$--5$_4$) blue- and redshifted
(continuum subtracted) emissions towards IRAS2A.
The images trace the clumps well at different velocities along
the N-S jet driven by MM1 and also trace the redshifted E 
lobe associated with MM2. The grey lines show the deceleration of the 
highest velocity clumps. 

Figure A.3 shows the SiO and SO PV diagrams along the E-W jet axis:  
as in the N-S case, the SiO emitting at the highest velocities is 
closely associated with the driving source MM2,
confirming that SiO is a powerful tracer of the jet launching region.

\begin{figure*}
\centerline{\includegraphics[angle=0,width=15cm]{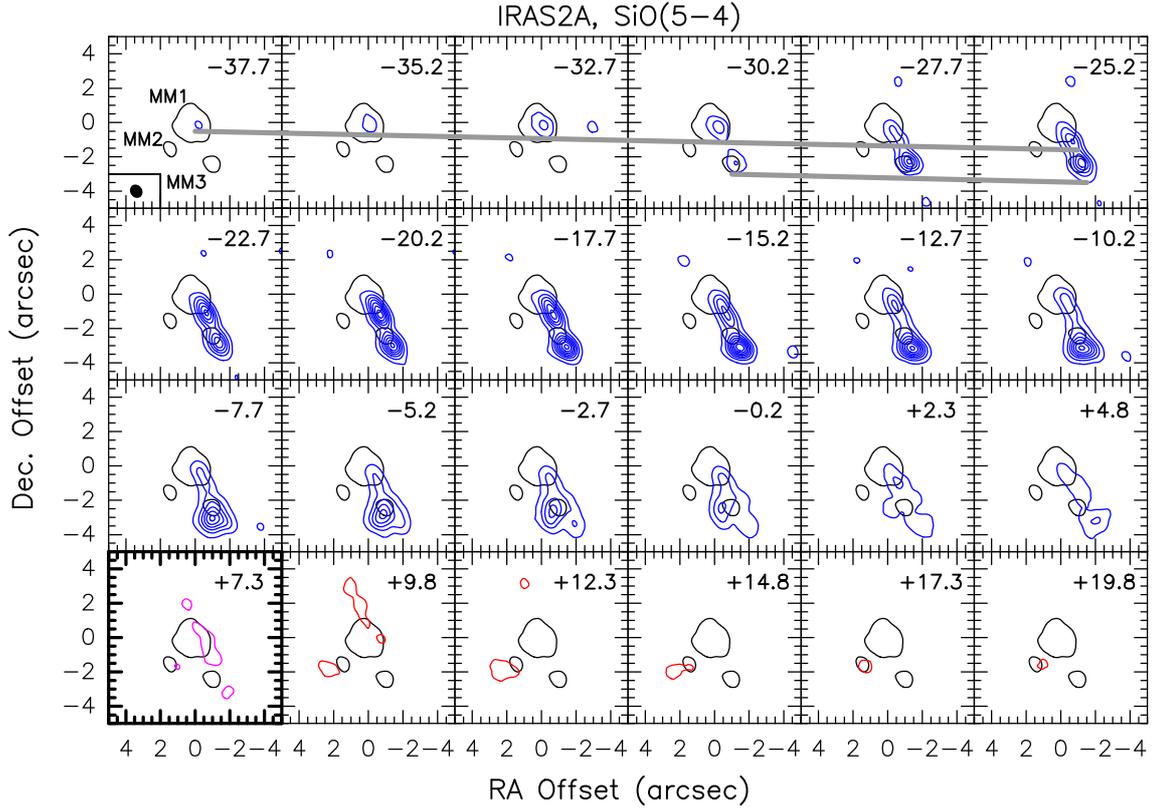}}
\caption{Channel maps of the SiO(5--4) blue- and redshifted
(continuum subtracted) emissions towards IRAS2A.
Each panel shows the emission integrated over a velocity interval
of 2.5 km s$^{-1}$ centred at the value given in the upper-right corner.
The thick box and the magenta contours indicate 
the range associated with the systemic velocity.
Thick contours correspond to the 5$\sigma$ emission of the
1.4 mm continuum map shown in Fig. 1 and indicate the position of
the MM1, MM2, and MM3 continuum sources. The ellipse in the top-left panel shows
the PdBI synthesised beam (HPBW): $0\farcs81\times0\farcs69$ (PA = 33$\degr$).
First contours and steps correspond to 5$\sigma$ (15 mJy beam $^{-1}$ km s$^{-1}$)
and 10$\sigma$, respectively. Grey lines indicate the slowing down of the highest velocity
SiO clumps (see text).}
\label{siochannels}
\end{figure*}

\begin{figure*}
\centerline{\includegraphics[angle=0,width=15cm]{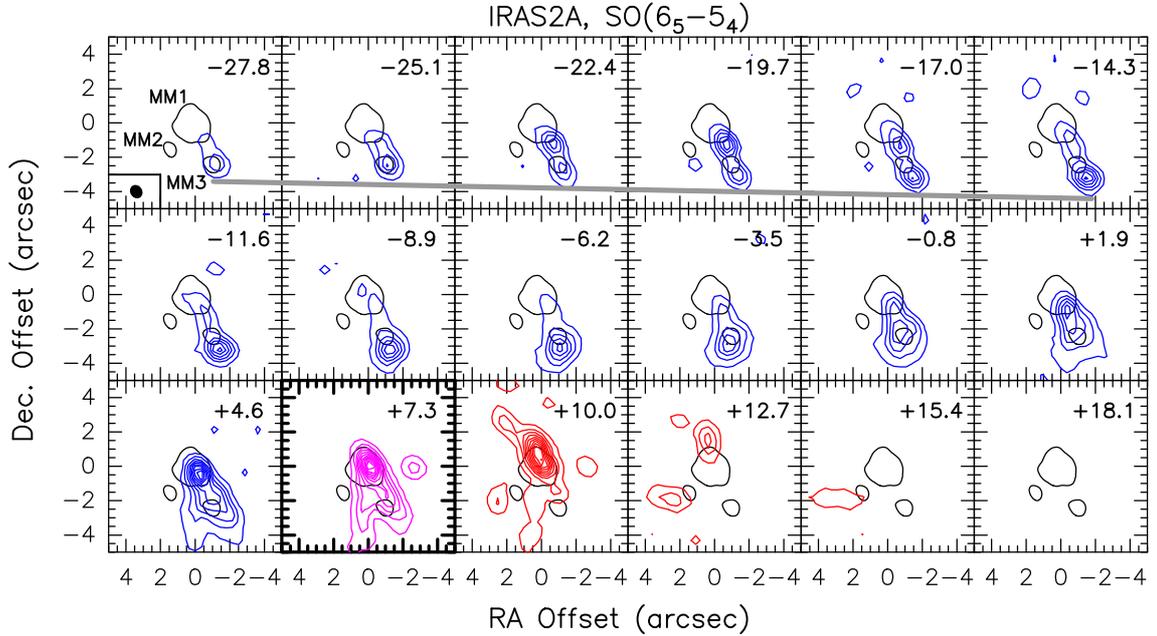}}
\caption{Channels map of the SO(6$_5$--5$_4$) blue- and redshifted
(continuum subtracted) emissions towards IRAS2A.
Each panel shows the emission integrated over a velocity interval
of 2.7 km s$^{-1}$ centred at the value given in the upper-right corner.
Symbols are drawn as in Fig. 2.
The ellipse in the top-left panel shows
the PdBI synthesised beam (HPBW): $0\farcs81\times0\farcs69$ (PA = 33$\degr$).
First contours and steps correspond to 5$\sigma$ (15 mJy beam $^{-1}$ km s$^{-1}$)
and 10$\sigma$, respectively.
Grey lines indicate the slowing down of the highest velocity
SO clump (see text).}
\label{sochannels}
\end{figure*}

\begin{figure}
\centerline{\includegraphics[angle=0,width=7.5cm]{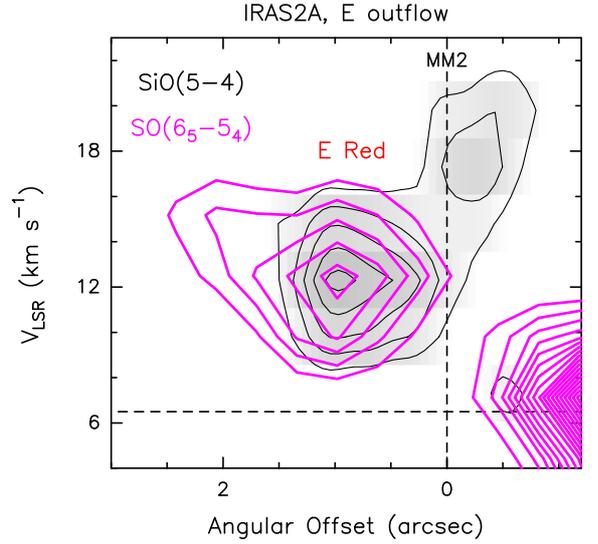}}
\caption{Position-velocity cut of SiO(5--4) (grey scale and black contours)
and SO(6$_5$--5$_4$) (magenta contours) along the
whole E-W jet (PA = 105$\degr$, see the grey line in Fig. 1).
First contours and steps correspond to 5$\sigma$ (2.5 K for SiO and 4.0 K for SO)
and 3$\sigma$, respectively.
Dashed lines mark the positions of MM2 and the
protostellar envelope
$V_{\rm LSR}$ (+6.5 km s$^{-1}$). We note that the SiO and SO emission at 
negative angular offsets traces the N-S outflow driven by MM1 (see Fig. 1).} 
\label{pvsioso2}
\end{figure}

\end{document}